\documentclass[twocolumn,groupedaddress,
superscriptaddress,preprintnumbers,
amsmath,amssymb,
aps,
]{revtex4}%-1}
\usepackage{graphicx}% Include figure files
\usepackage{dcolumn}% Align table columns on decimal point
\usepackage{bm}%
\usepackage[colorlinks=true,citecolor=blue]{hyperref}
\hypersetup{colorlinks=true,citecolor=blue,linkcolor=red,urlcolor=blue}

\usepackage{float}
\usepackage{amsmath}
\newcommand{\angstrom}{\mbox{\normalfont\AA}}
\usepackage{color}
\usepackage{xcolor}
\usepackage[colorlinks=true,citecolor=blue]{hyperref}
\hypersetup{colorlinks=true,citecolor=blue,linkcolor=red
,urlcolor=blue}

\usepackage{amsmath,amssymb}
\usepackage{color}
\usepackage{graphicx}
\usepackage{natbib}

\begin{document}
\title{$\sigma_h$ symmetry and electron-phonon interaction in two-dimensional crystalline systems}
\author{Mohammad Alidoosti}
\affiliation{Pasargad Institute for Advanced  Innovative Solutions (PIAIS), Tehran 19916-33361, Iran}
\author{Davoud Nasr Esfahani}
\email{d.nasr@khatam.ac.ir;dd.nasr@gmail.com}
\affiliation{Pasargad Institute for Advanced  Innovative Solutions (PIAIS), Tehran 19916-33361, Iran}
\affiliation{Department of converging technologies, Khatam University, Tehran 19916-33357, Iran}
\author{Reza Asgari}
\email{asgari@ipm.ir}
\affiliation{School of Physics, Institute for Research in Fundamental Sciences (IPM), Tehran 19395-5531, Iran}
\affiliation{School  of  Physics,  University  of  New  South  Wales,  Kensington,  NSW  2052,  Australia}
%\affiliation{ARC Centre of Excellence in Future Low-Energy Electronics Technologies, UNSW Node, Sydney 2052, Australia}
\date{\today}

\begin{abstract}
The coupling of electrons and phonons is governed wisely by the symmetry properties of the crystal structures. In particular, for two-dimensional (2D) systems, it has been suggested that the electrons do not couple to phonons with pure out-of-plane distortion, as long as there is a $\sigma_h$ symmetry. We show that such a statement is correct when constituents of the unit-cell layer are only located in the $\sigma_h$ symmetric plane; a prominent example of such a system is graphene. For those 2D crystals in which atoms are vertically located away from the horizontal symmetric plane (e.g., 1H transition metal dichalcogenides), acoustic flexural modes do not couple to the electrons up to linear order, while optical flexural phonons, which preserve $\sigma_h$ symmetry, do couple with the electrons.
Our conclusions are supported by an analytic argument together with numerical calculations using density functional perturbation theory.
\end{abstract}
\maketitle

\section{Introduction}
Atomically thin two-dimensional (2D) materials, including one- (several-) atomic-width layer(s), have sparked a great deal of attention owing to their various applications with nano-technological instruments~\cite{novoselov2005two,RevModPhys.82.2673,wang2012electronics,geim2013van}. There are a variety of crystals in the context of 2D systems that consist of atoms that do not lie in a single layer. Prominent examples are buckled structures such as silicene and germanene, a puckered structure such as black-phosphorene, and transition metal dichalcogenides (TMDs)\cite{PhysRevLett.102.236804,doi:10.1021/acs.jpclett.5b01094,KHANDELWAL201717,doi:10.1021/nl400516a,doi:10.1021/jp212558p}. In particular, $\sigma_h$ symmetry holds for 1H-TMDs 2D layers although atoms do not lie vertically in a single plane. This is equally true for other types of dichalcogenides, such as $\alpha$ phase 2D Iindium and gallium chalcogenides~\cite{C7NR09486H,doi:10.1021/nl4010089}.

In the long-wavelength limit, distortion in the out-of-plane direction of 2D materials results in the emergence of flexural modes (ZA and ZO referring to acoustic and optical out-of-plane phonon branches, respectively) in the phonon dispersion which could strongly affect the physical properties of materials, such as the mobility of a normal state, thermal and mechanical properties, superconductivity, charge density wave, and exciton-phonon interaction~\cite{Jiang_2015,grimvall1981electron,PhysRev.108.1175,peierls1996quantum,kohn1959image,PhysRevB.76.165108,calandra2010adiabatic,RevModPhys.77.1173,Zhu2367}.
The importance of the either existence or absence of mirror symmetry will be more obvious if the transport properties of materials are considered in the presence of phonons. The high carrier mobility in graphene is expected as long as mirror symmetry is supposed ~\cite{PhysRevLett.118.046601}, while the symmetry breaking (via an electrostatic gate, for instance) leads to a substantial reduction in the intrinsic mobility owing to the enhancement of ZA phonon scattering. Besides, acoustic flexural phonons lead to a remarkable lattice thermal conductivity of graphene~\cite{Jiang_2015,PhysRevB.82.115427}. Moreover, intrinsic carrier mobility is severely affected in the presence of such fluctuations in silicene~\cite{Rengel_2018}. For hexagonal group-V structures, Rudenko et al.~\cite{PhysRevB.100.075417} reported that acoustic out-of-plane deformations play a substantial role in restricting intrinsic mobility at low carrier concentrations. In fact, such a reduction arises from divergence of the ZA phonon population and it has already been mentioned by the Mermin-Wagner theory~\cite{PhysRevB.93.155413}.

Heretofore, theoretical studies have almost considered the effects of ZA phonons and less attention has been devoted to ZO phonon modes. On the other hand, the properties related to structures with broken $\sigma_h $ symmetry were well-inspected~\cite{PhysRevLett.126.117402, zhang2021role}. Therefore, the lack of a comprehensive investigation of the coupling between electrons and optical out-of-plane vibrations is more apparent than before, especially in the presence of $\sigma_h $ symmetry. In our paper, we show that stemming from both analytic analysis and density functional perturbation theory, the coupling of electrons and phonons with out-of-plane distortions (both ZA and ZO phonon modes) is different for two types of $\sigma_h$ symmetric crystals. For those systems whose atoms lie in the $\sigma_h$ plane, the flexural (optical and acoustic) phonons do not couple to electrons up to linear order \cite{doi:10.1021/nl402696q,PhysRevB.76.045430}. For those crystals whose atoms lie away from the $\sigma_h$ symmetric plane (such as 2D 1H-TMDs), only acoustic flexural phonons do not couple with the electrons, but, certain types of optical flexural vibrations do couple with the electrons. The former has already been proven for materials such as graphene with a pure 2D nature, i.e., the constituents of the crystal lie in exactly the same plane, and subsequently, the plane forms the center of reflection in the absence of any external perturbation.

In the context of the simulation of materials, state-of-the-art first-principles methods such as density functional perturbation theory (DFPT)~\cite{RevModPhys.73.515} or supercell methods can be used to calculate the magnitude of the electron-phonon couplings (EPCs). On an equal footing, symmetry analysis of the EPCs is still very instrumental for exploring the system. Therefore, we pursue these methods to perceive the coupling between itinerant charge carriers with phonons.  

Our paper is organized as follows. In Sec.~\ref{sec2} we provide our analysis corresponding to the coupling of electrons and flexural phonons for two various types of the above mentioned $\sigma_h$ symmetric systems. In Sec.~\ref{sec3} we provide some numerical examples of the electron-phonons interaction for realistic systems utilizing DFPT theory and finally, we present our conclusions in Sec.~\ref{sec4}.

%\newpage
\section{Symmetry analysis and flexural phonons}\label{sec2}
For the sake of simplicity, we consider a two-dimensional crystalline structure that consists of two similar atoms per unit-cell in a long-wavelength limit $\bf q\rightarrow 0$. To explore the dynamics of vibrations in a discrete lattice system, we usually assume the nuclear motion is slow so that the relatively light electrons are able to follow the nuclear motion adiabatically and for any configuration of the nuclei, the electrons are in the ground state. The Schr${\ddot{\text o}}$dinger equation for the electron wave function is defined by the effective potential energy, $V_{scf}({\bf r}) $\cite{Note1}, which implicitly depends on the configurations of nuclei~\cite{RevModPhys.73.515}. There are modes with pure out-of-plane distortion, and displacement vectors are given by $(u,\pm u)$ where $\pm$
stands for the acoustic (in-phase) and optical (out-of-phase) modes, respectively. In this manner, the deformation potential for acoustic and optical modes reads as follows: 
 \begin{eqnarray}\label{eq11}
  \Delta_{A}V_{scf}({\bf r})  =  \left ( \frac{\delta V_{scf}({\bf r}) }{\delta u_{1}} + \frac{\delta V_{scf}({\bf r}) }{\delta u_2}   \right ) u, \nonumber \\
  \Delta_{O}V_{scf}({\bf r})  = \left ( \frac{\delta V_{scf}({\bf r}) }{\delta u_1} - \frac{\delta V_{scf}({\bf r}) }{\delta u_2}   \right ) u, 
 \end{eqnarray}
where $\frac{\delta}{\delta u_i}$ corresponds to the variation of the potential associated with the displacement of
the $i$th atom in the $z$-direction, and $O$ and $A$ correspond to the optical and acoustic flexural modes, respectively. Notice that the $\bf q$ index is omitted for simplicity. Moreover, $V_{scf}({\bf r})$ represents the self-consistent potential, which is an even function under $\sigma_h$ operation~\cite{RevModPhys.73.515,Note1}. If we consider $\sigma_h$ symmetry, we will include two various configurations:
(a) Two atoms construct a planar monolayer such that two species lie in the same layer and the layer forms the center of $\sigma_h$ symmetry. In this case for each $i=1,2$ we obtain
   \begin {eqnarray}
    \frac{\delta  { V_{scf}}(x,y,-z)}{\delta u_i} =-\frac{\delta { V_{scf}}(x,y,z)}{\delta u_i},
   \end {eqnarray}
therefore,
\begin{eqnarray}
 \Delta_{A} { V_{scf}}(x,y,-z) &=&  \left ( \frac{\delta  { V_{scf}}(x,y,-z)}{\delta u_1} + \frac{\delta  { V_{scf}}(x,y,-z)}{\delta u_2}   \right ) u  \nonumber \\
 &=& \left ( -\frac{\delta  { V_{scf}}(x,y,z)}{\delta u_1} - \frac{\delta { V_{scf}}(x,y,z)}{\delta u_2}   \right ) u \nonumber \\
 &=& - \Delta_{A} { V_{scf}}(x,y,z)
\end{eqnarray}
$\Delta_{A} { V_{scf}}(x,y,z)$ is an odd function under $\sigma_h$ operation. By making use of similar arguments, one infers the same conclusion for the optical flexural mode, i.e., $\Delta_{O} { V_{scf}}(x,y,-z) = -\Delta_{O}{ V_{scf}}(x,y,z)$.
Since $\Delta V_{O/A}$ is an odd function under $\sigma_h$ operation and because $\psi_k(r)$ is either even or odd under $\sigma_h$, the matrix elements $\langle \psi_k|\Delta_{A,O}V|\psi_k\rangle$ become zero. Accordingly, an intraband transition owing to flexural phonon perturbation is no longer allowed for the atomic configurations in which atomic species lie on the horizontal plane.
(b) Two atoms do not lie on the horizontal mirror plane, yet $\sigma_h$ symmetry persists. In this case, the following relation holds for $i,j=1,2$ and $i\neq j$:
 \begin {eqnarray}
    \frac{\delta  { V_{scf}}(x,y,-z)}{\delta u_i} =-\frac{\delta  { V_{scf}}(x,y,z)}{\delta u_j}.
   \end {eqnarray}
   
Using this equation, we thus have
\begin{eqnarray}
 \Delta_{A} { V_{scf}}(x,y,-z) &=&  \left ( \frac{\delta  { V_{scf}}(x,y,-z)}{\delta u_1} + \frac{\delta  { V_{scf}}(x,y,-z)}{\delta u_2}   \right ) u  \nonumber \\
 &=& \left ( -\frac{\delta  { V_{scf}}(x,y,z)}{\delta u_2} - \frac{\delta  { V_{scf}}(x,y,z)}{\delta u_1}   \right ) u. \nonumber \\
 &=& - \Delta_{A}{ V_{scf}}(x,y,z),
\end{eqnarray}

The deformation potential due to the acoustic phonons is an odd function with respect to $z$.
However, for the optical phonons, we have
\begin{eqnarray}
 \Delta_{O} { V_{scf}}(x,y,-z) &=&  \left ( \frac{\delta { V_{scf}}(x,y,-z)}{\delta u_1} - \frac{\delta  { V_{scf}}(x,y,-z)}{\delta u_2}   \right ) u  \nonumber \\
 &=& \left ( -\frac{\delta { V_{scf}}(x,y,z)}{\delta u_2} + \frac{\delta  { V_{scf}}(x,y,z)}{\delta u_1}   \right ) u \nonumber \\
 &=& \Delta_{O} { V_{scf}}(x,y,z),
\end{eqnarray}
which is an even function under $\sigma_h$ operation.
Therefore, the intraband transitions are forbidden for the acoustic phonons, while the matrix elements $\langle \psi_k |\Delta_O V|\psi_k\rangle $ generally are unequal to zero for the optical phonons based on symmetry considerations.
Accordingly, $\sigma_h$ symmetry is an insufficient condition to decouple flexural phonons from the electrons for certain optical phonons (which maintain $\sigma_h$ symmetry).

%%%%%%%%%%%%%% FIRST PRINCIPLES

\begin{figure*}[t]
\centering
\includegraphics[scale=0.50,trim=1.4cm 1cm 0 0]{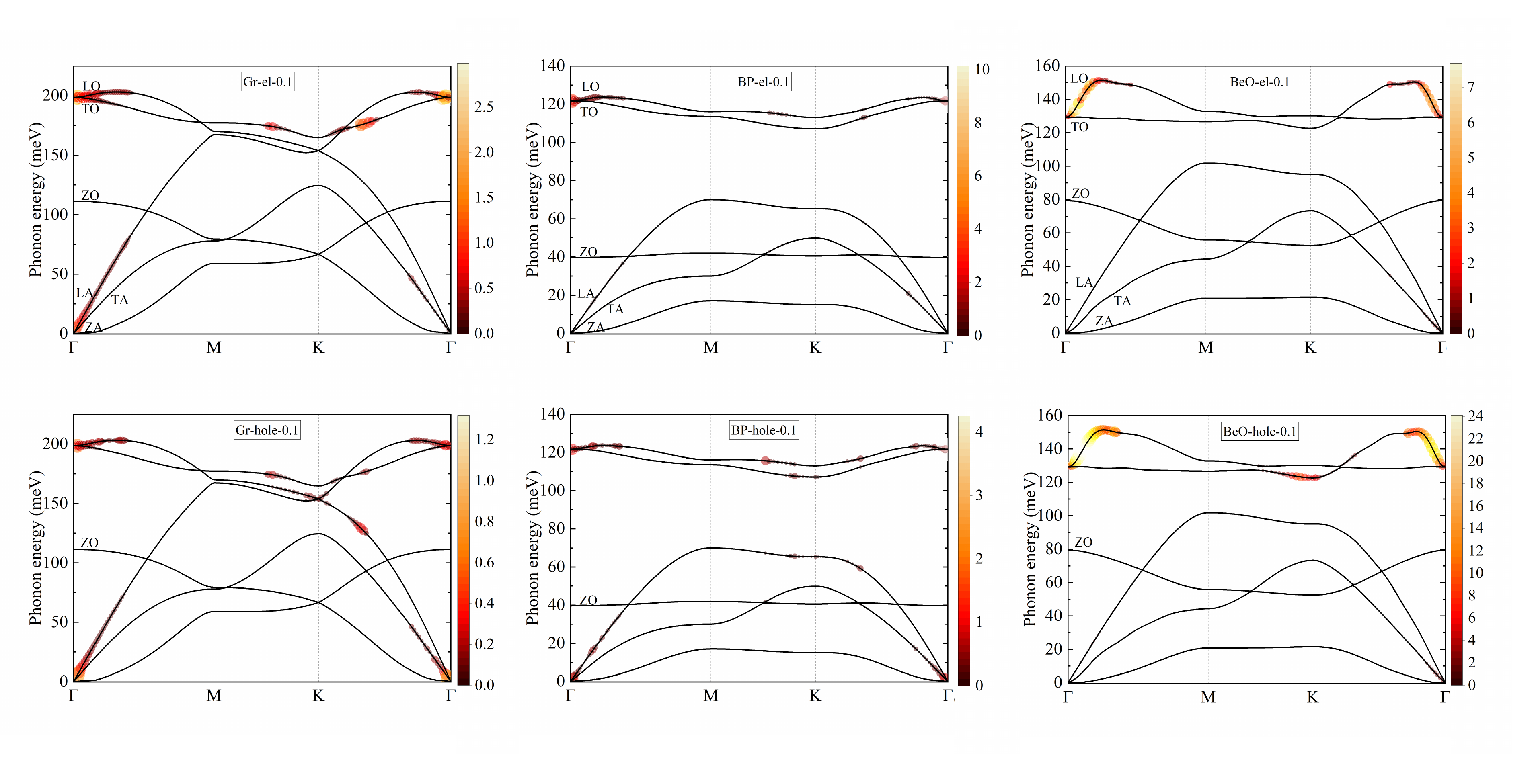}
\caption{(Color online) The strength of the electron-phonon interaction for various modes of the phonon band along the high symmetry path for the electron and hole doping (0.1) of monolayer graphene (left panels), BP (middle panels), and BeO (right panels). The ZO label represents the optical out-of-plane deformations. There is no coupling between the electrons and flexural phonons for both the electron and hole-doped cases and in the ZA and ZO displacements in the BP and Gr cases. However, the LO branch receives the most enormous contribution of the EPC in the BeO structure. Therefore, the flexural modes will not participate in the formation of $\lambda_{{\bf q}\nu}$.}
\label{P2D-ph}
\end{figure*}

\section {Numerical examples: DFPT simulations}\label{sec3}
First-principles calculations are performed within the density functional theory (DFT) framework implemented in a \textsc{Quantum Espresso} distribution~\cite{0953-8984-21-39-395502} which works based on a pseudopotential method. Our calculations are carried out by making use of the norm-conserving pseudopotential and choosing the local density approximation (LDA) as the exchange-correlation functional. In addition, the calculations are relaxed by using the Broyden–Fletcher–Goldfarb–Shanno algorithm (BFGS) until all forces and the pressures are less than $ 10^{-5} \, eV/\angstrom $ and 0.5 kbar, respectively. The Kohn-Sham wave functions are cut at $\sim 100$ $Ry$. In addition, periodic boundary conditions are applied along the $x$ and $y$ directions, while, a vacuum space of 20 $\angstrom$ along the $z$ direction is adopted to wipe out the spurious interaction. The DFPT is employed to compute the phonon modes and the EPC matrix elements of the system \cite{baroni2001phonons}. To increase accuracy and efficiency in the electron-phonon calculations and associated properties, a fine sampling of the electron and phonon wavevectors within the Brillouin zone is necessary. Therefore, maximally localized Wannier functions \cite{marzari1997n,souza2001maximally,mostofi2008wannier90} are used; implemented within the \textsc{EPW} code\cite{ponce2016epw}. Details of coarse and fine $\mathbf k$- and $ \mathbf q$-points for various structures  are listed in Ref.~\cite{kqpoints}. Moreover,  the  $ \mathbf k$-point grids   applied for electronic integration with a Monkhorst-Pack mesh \cite{PhysRevB.13.5188} are similar to the coarse grids~\cite{kqpoints}.
%A coarse grid $12\times 12\times1$ is used for both $ \mathbf k$- and $ \mathbf q$-points to calculate the electronic band structure and phonon dispersion and a fine grid $300\times 300\times1$ $ \mathbf k$-points and $150\times 150\times1$ $ \mathbf q$-points are used to evaluate the electron-phonon related properties.

To represent the EPC strength, we calculate $ \lambda_{\bf q, \nu}$, as a dimensionless value associated with a single phonon momentum $\bf q $ and mode $\nu $ and it can be written as~\cite{giustino2017electron,PhysRevB.103.035411}:
\begin{equation}
\begin{split}
\lambda_{{\bf q}\nu} = 
\frac{1}{N_0\omega_{{\bf q}\nu}}\sum_{mn,{\bf k}}\mathrm{w}_{{\bf k}} |g_{mn,\nu}({\bf k,q})|^2 \\ \times \delta(\varepsilon_{n{\bf k}}-\varepsilon_F)\delta(\varepsilon_{m{\bf k}+{\bf q}}-\varepsilon_F),
\end{split}
\label{eph-strength}
\end{equation}	
where $N_0$ is the density of states at the Fermi level ($\varepsilon_F$), $ \mathrm{w}_{{\bf k}}$ is the weight of $ \mathbf k$-points in whole Brillouin zone and $\varepsilon_{n{\bf k}} $ and $\varepsilon_{m{\bf k}+{\bf q}}$ are Kohn-Sham eigenvalues obtaining from unperturbed DFT. The $ g_{mn,\nu}({\bf k,q})$ is the EPC matrix elements specifying a scattered electron from state n$\bf k$ to m$\bf k + q$ where $n$ and $m$ are the electronic band number with momentum $\bf k $ and $\bf k ^ \prime=\bf k + q$, via absorbing or emitting a phonon with momentum $\bf q$, mode $ \nu$ and frequency $ \omega$.
In practice, we model the $\delta-$function of the above integrand as a Gaussian function with a small width. To do so, we apply width broadening, $\sigma_{el} = 10\ \text{meV}$ for electronic states %- and $\sigma_{hole} = 0.2 \ \text{meV}$ for hole-doping 
accompanied by a fine mesh grid as mentioned above that guarantees the convergence of $\lambda$ until it becames broadening-independent.
Our numerical results show that the EPC disperses slightly in terms of $\sigma_{el}$. %This manner provides an appropriate EPC showing an accurate behavior of the phonon modes contribution. 
  \begin{figure*}[t]
\centering
\includegraphics[scale=0.5,trim=1.4cm 1cm 0 0]{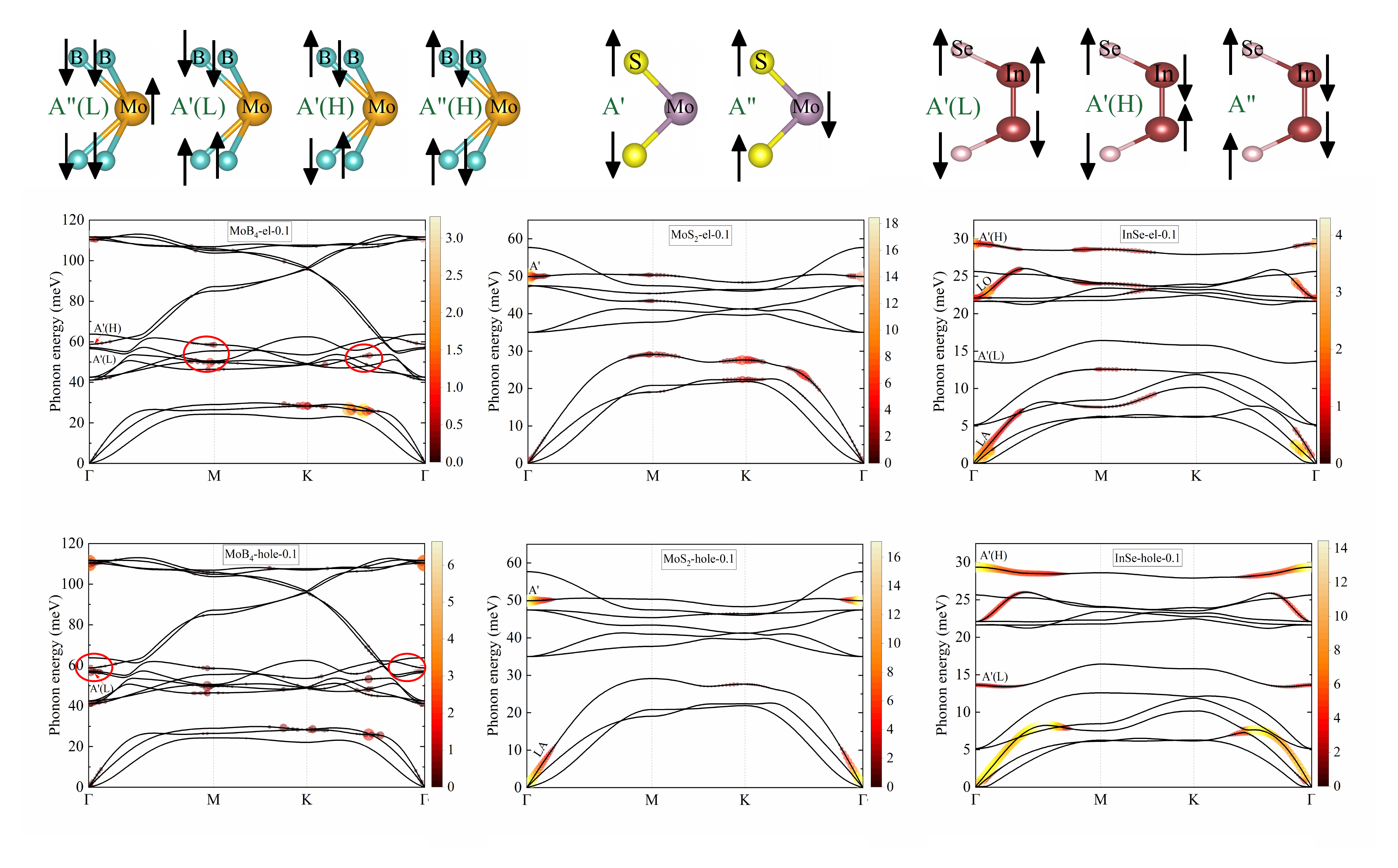}
\caption{(Color online) The momentum resolved EPC strength, $\lambda_{{\bf q}\nu}$, mapped on the phonon dispersion along the high symmetry path for the electron and hole doping (0.1) of monolayer MoB$_4$ (left panels), MoS$_2$(middle panels) and InSe (right panels). The out-of-plane deformation of the optical branch is displayed by A$'$ label. The most determined EPC strength occurred by the ZO deformations for $\textbf q \rightarrow 0$ in MoB$_4$ and red elliptical lines are used to clarify the contribution ZO during the EPC process. In addition, the importance of the ZO polarization could be increasingly evident in monolayer III-VI post-transition metal chalcogenides. Symmetry representations related to the out-of-plane optical displacement of quasi 2D crystals at the zone center are depicted in the first row. Labels L and H refer to low and high phonon frequencies.}
\label{Q2D-ph}
\end{figure*}
Two types of structural phases are considered; purely 2D structures (P2D) namely, graphene (Gr), boron phosphide (BP) and beryllium oxide (BeO), and quasi 2D (Q2D) structures, such as molybdenum tetraboride (MoB$_4$), MoS$_2$ and InSe
 all of which possess a hexagonal structure and the $\sigma_h$ symmetry is preserved for all structures studied here.

 Our relaxed geometry calculations for these materials are in good agreement with those obtained by others~\cite{PhysRevB.80.155453,doi:10.1021/acs.jpcc.6b06161}. The density of states (DOS) must have a finite value to calculate $\lambda_{{\bf q}\nu}$, therefore, we consider the impact of doping level 0.1 ($e/{\text {unitcell}}$) for both of the electron/hole cases through a rigid raising or lowering of the Fermi level from the conduction band minimum (CBM) or the valance band maximum (VBM), respectively. For simplicity, we drop ($e/{\text {unitcell}}$) unit in the doped cases.
To determine the influence of the $\lambda_{{\bf q}\nu}$ related to two specific branches, ZA and ZO, the phonon band structure of the monolayer P2Ds is depicted in Fig.~\ref{P2D-ph} for both the electron and hole doping 0.1 along with the high symmetry directions of the first Brillouin zone (1BZ). The colors represent the strength of the EPC, while it is integrated over the whole of the electronic momentum $\textbf k$. In addition, we set labels related to the in-plane modes with longitudinal acoustic (LA) and transverse acoustic (TA) displacements around the $\Gamma$ point. The next two in-plane branches include the transverse optical (TO) and longitudinal optical (LO) modes.
In the case of doped graphene, the left panels of Fig.~\ref{P2D-ph} show that there is no coupling between the electrons and flexural phonons including both the ZA and ZO displacements for both
the electron- and hole-doped cases. More details reveal that bright spots located at the fourth branch at the middle of the K-$\Gamma$ path for hole-doped graphene are related to the LA atomic displacements. It should be noticed that the magnitudes of the longitudinal modes, $\lambda_{{\bf q}\nu}$ are comparable for electron- and hole-doped graphene so that it conducts to a total $\lambda$ of about 0.19 and 0.07, respectively, which agrees well with those results reported in Ref.~\cite{PhysRevB.90.014518}.
The ZO and ZA phonon modes do not have a contribution in the EPC for BP, like Gr, as depicted in the middle panels of Fig.~\ref{P2D-ph}. In this case, the LA vibrations primarily result in a tangible amount of the EPC strength only close to the zone center and other phonon wave vectors ($\textbf q$) have a marginal role, especially in the electron-doped case. Although the maximum of the EPC strength in BP is greater (by factor 3) than that of graphene, the total $\lambda$ adopts values 0.14 and 0.07 for the electron and hole-doped BP due to the low number of active ${\bf q}$’s during the scattering process.
The right panels of Fig.~\ref{P2D-ph} describe the EPC strength mapped on the phonon dispersion for both the electron- and hole-doped BeO along with all high-symmetry directions. While there are no ZO and ZA phonon contributions, the LO phonon branch has the most enormous contribution of the EPC according to what has been underlined by Ref.~\cite{doi:10.1063/5.0022426} so that for the electron doped case (right-top panel) due to the presence of a single sheet at the $\Gamma$ point (see Appendix; Fig.~\ref{BeO-FS}),  a remarkable amount of LO phonon scattering occurs only in the vicinity of the zone center. In the case of electron-doped BeO, inherent mobility may still be limited, while the hole-doped case (bottom-right panel) suggests the LO phonon modes are expected to be strongly coupled with electron states not only for $\textbf q \simeq 0$, but also for $\textbf q \simeq \textbf K$; including the LO phonon polarization. More details reveal that the phonon wave vector $\textbf q \simeq \textbf K$ is related to the inter-pocket scattering of the electrons located at the $\textbf K'_2$ and $\textbf K_2$ points (also phonons with a $\frac{2\pi}{3}$ rotation of $\bf q\simeq \textbf K$ around the zone center) at the Fermi sheet in the 1BZ, see Appendix, Fig.~\ref{BeO-FS}. On the other hand, $\lambda_{\bf q}$ with $\bf q\rightarrow 0$ arises from the intrapocket scattering processes. Moreover, a comparison between the electron- and hole-doped cases shows that the maximum strength of the EPC in the hole-doped case is considerably greater than that of the electron-doped case, which is due to the larger density of states in the hole-doped regime (see the Appendix). 

 To sum up,  as long as P2D structures are protected by mirror symmetry, the flexural modes will not participate in the formation of $\lambda_{{\bf q}\nu}$ and significantly the longitudinal deformations play a crucial role in the EPC processes. This behavior is more obvious in the long-wavelength limit, which is compatible with other results linked to the EPC strength~\cite{he2020investigation,PhysRevB.85.165440,PhysRevLett.105.266601,PhysRevB.77.115449,PhysRevB.93.035414}. 
 %
%%%%%%%%%%%%%%%%% Q2D %%%%%%%%%%%%%%%%%%%%%%%%%%%%%%%%%%%%%%%
Now, we investigate other examples beyond purely 2D crystals with several atomic layers; Q2D materials are constructed by several layers where the $\sigma_h$ symmetry is still present. In particular, Fig.~\ref{Q2D-ph} shows the strength of the EPC for Q2D structures, MoB$_4$, MoS$_2$, and InSe, as mapped onto the phonon dispersion along with the high symmetry points for both the electron- and hole-doping 0.1. In the phonon modes, the letter A$'$ refers to the symmetric representation corresponding to the out-of-plane optical displacements of Q2D crystals as long as $\sigma_h$ symmetry is retained. These modes are expected to couple to the electrons following our discussions in Sec.~\ref{sec2} and they will be called breathing [see Fig.~\ref{Q2D-ph}]. On the other hand, A$''$ modes are not $\sigma_h$ symmetric and the atoms which are equivalent under $\sigma_h$ have in-phase deformations. Consequently, it is straightforward to check that the deformation potential corresponding to A$''$ modes is still odd under $\sigma_h$ reflection. Accordingly, they do not couple to electrons, similar to the ZA phonon modes.
The left panels of Fig.~\ref{Q2D-ph} show that the most effective phonon contribution is related to $\lambda_{{\bf q}\nu}$ for the electron-doped MoB$_4$, which originates from the in-plane optical modes (the highest branch) at the zone center. The acoustic in-plane vibrations accompanied with the ZO-A$'$ phonon modes (labeled by red elliptical lines and far away from the zone center) have sensitive amount of $\lambda_{{\bf q}\nu}$, while there is no contribution from the ZA vibrations during the EPC process.

In the case of hole-doped MoB$_4$, the $\lambda_{{\bf q}\nu}$ vanishes for the ZA phonon modes along the whole $\textbf q$ in the high symmetry path. On the other hand, the strongest EPC strength occurs by the ZO-A$'$(L) deformations for $\textbf q \rightarrow 0$ as represented by red elliptical lines. Such out-of-plane displacement, i. e., the Raman active displacement, is illustrated on the top of the first column
of Fig.~\ref{Q2D-ph}. In addition, the ZO-A$'$ phonons have a notable impact on the electron-phonon scattering rate and could strongly restrict carrier mobility. Notice that the momentum resolved EPC strength, $\lambda_{{\bf q}\nu}$, in MoB$_4$ is comparable with that related to the graphene result, so that a total $\lambda$ of about 0.14 and 0.17 is calculated for the electron- and hole-doping, respectively, implying high enough mobility for both the electron and hole carriers.  
The middle panel of Fig.~\ref{Q2D-ph} shows that the ZO-A$'$ and LA phonon branches mainly contribute to the EPC strength for 1H-MoS$_2$ at $\textbf q \rightarrow 0$ for both the electron- and hole-doped cases. Representation of these optical out-of-plane deformations is specified by the A$'$ label in Fig.~\ref{Q2D-ph}, where only ZO-A$'$ phonon modes with a breathing vibration couple to electrons. Moreover, there is a specific $\textbf q \simeq K$ whose inter-valley scattering affected electronic transport properties as reported in Refs.~\cite{PhysRevMaterials.2.114010,PhysRevB.85.115317}. 

The total $\lambda$ are evaluated with the values of 0.82 and 0.73 for the electron- and hole-doping, respectively. In addition, the ZA distortion does not have any contribution in the formation of the $\lambda_{{\bf q}\nu}$ for both doped cases.
It is worth mentioning that the importance of the polarization ZO-A$'$ phonon mode could be increasingly evident in monolayer III-VI transition metal chalcogenides (e.g., InSe) presented in the right panels of Fig.~\ref{Q2D-ph}. The ZA mode does not contribute to the electron-phonon coupling. The most influential branches are four (five) modes TA, LA, LO, and one (two) ZO mode(s) for electron- (hole-) doped levels. For optical out-of-plane displacements, sixth and twelfth phonon branches are depicted by A$'$(L) and A$'$(H) labels, respectively. More details reveal that both A$'$(L) (with an extremely limited amount for the electron-doped case) and A$'$(H) branches have contributions in the EPC for doping systems. The other ZO modes which do not preserve $\sigma_h$ symmetry and are illustrated by A$''$ labels in the first row of the figure, do not contribute to the EPC; their magnitude is exactly zero. Though the number of $\textbf q$'s that are active in the scattering process is large, we obtain a total $\lambda$ of about 0.6 in the electron-doped case, while the large number of active $\textbf q$'s accompanied with a remarkable amount of the EPC strength give rise to a total $\lambda$ $\sim 8$ in the hole-doped case.  The reason for the difference is the fact that the DOS for the hole-doped case is appreciably larger than that in the electron-doped case~\cite{PhysRevB.103.035411}.
Moreover, the discrepancy between the EPC values can be understood based on the density of states, the $g_{mn,\nu}({\bf q},\omega)$, and various lattice dielectric screenings of  TMDCs~\cite{torbatian2018plasmonic,felipe2020universal}. We would like to focus on the lattice dielectric screening to figure out its role in the EPC. In ordinary three-dimensional materials, the effect of the lattice screening is simply re-scaling of the interaction strength by a dielectric constant. In Q2D crystalline structures, however, the interaction is modified to $2\pi e^2/\epsilon(q+\alpha q^2)$, where $\epsilon$ is an averaged environment dielectric constant and $\alpha$ is related to the polarizability of the crystalline material and is defined as $\alpha=d \epsilon_{\parallel}/2.0$, where $d$ is a slab thickness and the isotropic in-plane dielectric constant is given by $\epsilon_{\parallel}$. To provide some numbers, $\epsilon_{\parallel} = 3.18$ and $5.13$
, and $d= 5.32$ and $3.13\angstrom$ for InSe and MoS$_2$, respectively. Therefore, $\alpha$ is $8.46$ and $8.03$ for those systems, respectively. This indicates that the lattice screening in InSe is greater than that in MoS$_2$ and thus the electron Coulomb interaction in InSe is smaller. Therefore,  we should conclude that the electronic Coulomb interaction is suppressed in InSe, and subsequently the EPC in MoS$_2$ is greater than in the InSe and MoB$_4$ samples.

%
%%%%%%%%%%%%%%%%%%%%%%%%%%  Conlusions

\section{Conclusions}\label{sec4}
In this work, we have investigated the electron-phonon coupling in 2D crystalline structures consisting of the $\sigma_h$ symmetric plane for both the ZA and ZO deformations. In order to perceive the electron-phonon coupling, symmetry consideration and first-principles calculations have been utilized. We have determined that there is no longer coupling between electrons with the ZA and ZO phonon modes for one-atomic 2D layer systems such as graphene, BP, and BeO systems. On the other hand, for a few atomic layers like MoB$_4$, MoS$_2$, and InSe, the ZA phonon modes do not couple to the electrons up to linear order, however, the optical flexural phonons encompassing A$'$ character could couple with the electrons. Accordingly, the behavior of these two branches will be completely distinct for pure and quasi 2D crystalline structures as long as $\sigma_h$ symmetry is preserved.
Therefore, the interaction of electrons with the ZO phonon modes within quasi 2D materials including $\sigma_h$ symmetry could affect various transport properties related to the electron-phonon couplings like carrier mobility and superconducting transition temperature.

\begin{acknowledgments}
 R. A. gratefully acknowledges the support from the Australian Research Council Centre of Excellence in Future Low-Energy Electronics Technologies (Project number CE170100039).
\end{acknowledgments}

\appendix
\section{}\label{app-a}
\begin{figure}[H]
\centering
\includegraphics[scale=0.29,trim={1cm 1cm 1cm 0}]{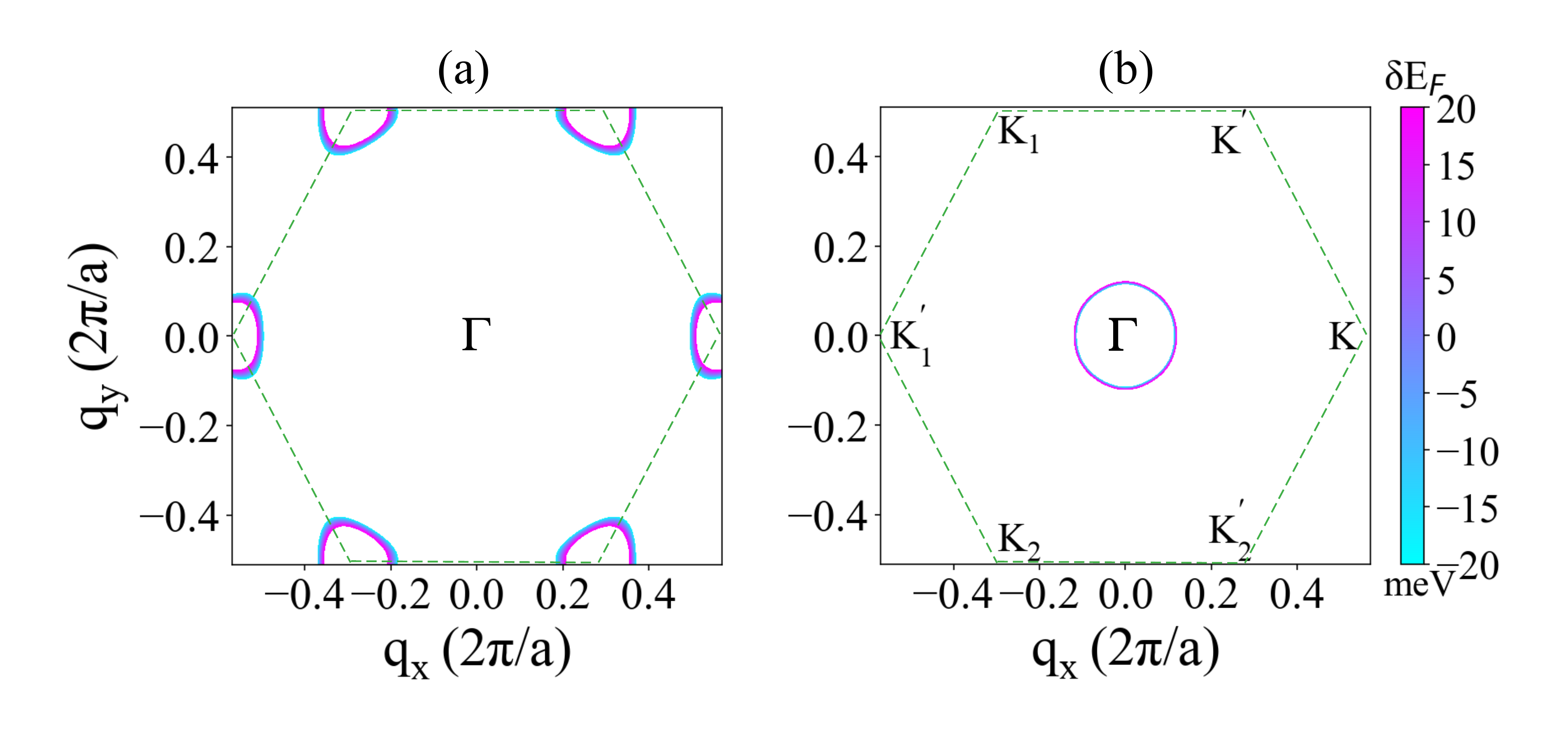}
\caption{(Color online) The Fermi surface contour of the monolayer BeO based on the rigid band model. The Fermi surface for doping (a) +0.1 and (b) -0.1.  The color bar shows the shift of the Fermi energy. The green dashed lines are applied to illustrate the first Brillouin zone boundaries.}
\label{BeO-FS}
\end{figure}

For electron and hole doping, the Fermi surface of monolayer BeO is illustrated in Fig.~\ref{BeO-FS}.
For example +0.1, Fig.~\ref{BeO-FS}(a) shows the presence of six semi-triangle pockets surrounding K and K$^\prime$ sites, allowing interpocket scattering of states for phonons with $|\bf q| \simeq K$. 

The Fermi surface for the electron doping regime, shown in Fig.~\ref{BeO-FS}(b), is a single sheet at the $\Gamma$ point, permitting only state scatterings for $\bf q\rightarrow 0$. DOS and g (the el-ph interaction matrix elements) are two important variables in determining $\lambda $, according to \cite{giustino2017electron}.
Within the isotropic regime, $\lambda$ has a linear relationship with DOS at the Fermi levels, as defined by $\lambda =  \frac{ 2 N(E_F)\langle|g^2|\rangle }{\omega_0}$, where $N(E_F)$ is DOS at $E_F$ and $\langle|g^2|\rangle$ is average of the square of electron-phonon interaction.
As a result, the higher DOS at the Fermi level leads to the higher $\lambda$.
In the case of BeO, a higher DOS in the hole doping situation leads to a higher $\lambda$ than in the electron doping case (see table.~\ref{tab1}).

\begin{table}[H]
    \centering
    \caption{Density of  states (DOS) and $\lambda$ for BeO and InSe for two +0.1 (hole) and -0.1 (electron) doping levels.} 
\begin{tabular}{lcccc}
\hline \hline 
 & \multicolumn{2}{c}{\bf DOS (states/spin/eV/Unit-Cell)} & \multicolumn{2}{c} { $\lambda$ }  \\
 \cline{2-3}  \cline{4-5} 
   & \textbf{+0.1} &  \textbf{-0.1}  & \textbf{+0.1} &  \textbf{-0.1} \\
\hline

BeO  & 0.53  & 0.11 & 1.87  & 0.26\\ 

InSe  & 6.72   & 0.76 & 8 & 0.6 \\ 
\hline \hline
\end{tabular} 
\label{tab1}
\end{table}
\indent For BeO, $\frac{\lambda_{hole}}{\lambda_{electron}}\simeq 7$, on the other hand $\frac{N_{hole}(E_F)}{N_{electron}(E_F)}\simeq 5$, which indicates the enhancement of $\langle|g^2|\rangle$ for hole-doped case. Such enhancement of $\langle|g^2|\rangle$ could be attributed to inter-valley scattering, which was discussed in the main text. That is, shift of $E_F$ into the conduction band results in a Fermi surface which contains a single valley at $\Gamma$, resulting in only intrapocket scattering processes, whereas, the shift of $E_F$ into the valance band produces a Fermi surface which is composed of six semi-triangular pockets at K point (see Fig.~\ref{BeO-FS}), that leads to both inter-pocket and intra-pocket scattering processes, which eventually results in a larger average of electron-phonon interaction $\langle|g^2|\rangle$ for hole-doped case in comparison with electron one. A similar analysis is also valid for InSe (please see more details in Ref.~\cite{PhysRevB.103.035411}.\\

\nocite{apsrev41Control}
\bibliographystyle{apsrev4-1}
\bibliography{draft1.bib}

\begin{thebibliography}{53}
\bibitem [{kqp()}]{kqpoints}%
  \BibitemOpen
  \href@noop {} {}\bibinfo {note} {The coarse $ \mathbf k$ ($ \mathbf q$)-grids
  are considered tantamount to $24\times 24\times1$ ($12\times 12\times1$) for
  graphene; $12\times 12\times1$ ($12\times 12\times1$) for both BP and BeO;
  $10\times 10\times1$ ($10\times 10\times1$) for MoS$_2$; $12\times 12\times1$
  ($6\times 6\times1$) for both MoB$_4$ and InSe. While, the related fine grids
  are $300\times 300\times1$ ($150\times 150\times1$) for graphene, BP, BeO and
  MoS$_2$; $240\times 240\times1$ ($120\times 120\times1$) for both MoB$_4$ and
  InSe.}\BibitemShut {Stop}%
\end{thebibliography}

\end{document}